\documentclass[aps,pra,twocolumn,superscriptaddress]{revtex4-2}

\usepackage{amssymb}
\usepackage{amsmath}
\usepackage{xcolor}
\usepackage{braket}
\usepackage{kotex}

\usepackage{graphicx}
\usepackage{epstopdf}
\usepackage{dcolumn}
\usepackage{bm}
\usepackage{lipsum}
\usepackage{color}
\usepackage{ulem}
\usepackage[colorlinks=true,linkcolor=blue,citecolor=blue,urlcolor=blue]{hyperref}

\newcommand{\figref}[2][]{%
  \hyperref[#2]{\ref*{#2}\if\relax\detokenize{#1}\relax\else(#1)\fi}%
}

\begin{document}

\title{Dynamic similarity of vortex shedding in a superfluid flowing past a penetrable obstacle}

\author{Junhwan Kwon}
\affiliation{Department of Physics and Astronomy, Seoul National University, Seoul 08826, Korea}
\affiliation{NextQuantum, Seoul National University, Seoul 08826, Korea}

\author{Yong-il~Shin}
\email{yishin@snu.ac.kr}
\affiliation{Department of Physics and Astronomy, Seoul National University, Seoul 08826, Korea}
\affiliation{NextQuantum, Seoul National University, Seoul 08826, Korea}
\affiliation{Institute of Applied Physics, Seoul National University, Seoul 08826, Korea}


\begin{abstract}
We numerically investigate wake dynamics in a superfluid flowing past a penetrable obstacle.
The penetrable obstacle does not fully deplete the density, and we define an effective diameter $D_{\rm eff}$ from the Mach-1 contour of the time-averaged irrotational flow around the obstacle, which delineates the local supersonic region where quantized vortices nucleate. 
Using this flow-defined length, we construct a superfluid Reynolds number $\mathrm{Re}_{\rm s} \equiv (v_0 - v_c) D_{\rm eff}/(\hbar/m)$ with $v_0$ being the flow speed, $v_c$ as the critical velocity and $m$ as the particle mass, and show that  $\mathrm{Re}_{\rm s}$ organizes the wake dynamics across obstacle sizes and strengths: the transition from dipole-row emission to alternating vortex cluster shedding occurs for $\mathrm{Re}_{\rm s} \approx 2$ and the Strouhal number and drag coefficient collapse onto universal curves versus $\mathrm{Re}_{\rm s}$. 
These results extend the notion of dynamic similarity in superfluid flows to penetrable obstacles and show that the dynamically relevant length scale is determined by the supersonic region rather than by the geometric obstacle size.
\end{abstract}

\maketitle

\section{\label{sec:level1}Introduction}
Dynamic similarity refers to the principle that flows, with different absolute scales of length, time, and mass in fluid properties, can exhibit identical behavior when expressed in terms of appropriate dimensionless parameters~\cite{batchelor_2000}. 
A paradigmatic example is a flow past bluff bodies, in which wake transitions and forces are organized according to the Reynolds number $\mathrm{Re}=UD/\nu$~\cite{roshko_1954}, where $U$ is the flow velocity, $D$ is the characteristic length of the body and $\nu$ is the kinematic viscosity. 
Above $\mathrm{Re}\approx 50$~\cite{provansal_J.FluidMech._1987}, a periodic pattern of alternating vortices, known as the von K\'arm\'an vortex street, starts to form, 
and for larger values of $\mathrm{Re}$, the wake gradually loses strict periodicity, entering turbulent vortex shedding around $\mathrm{Re}=250$~\cite{Lienhard_1966}. 
Nevertheless, the lift force, normal to the direction of the background flow, still has a dominant frequency.
The vortex-shedding frequency $f_s$ is nondimensionalized through the Strouhal number $\mathrm{St}=f_\mathrm{s}D/U$, showing a universal $\mathrm{St}\text{--}\mathrm{Re}$ relation over a broad range of $\mathrm{Re}$, with $\mathrm{St}\simeq 0.2$ maintained up to $\mathrm{Re}\sim 10^{5}$~\cite{roshko_1954,roshko_J.FluidMech._1961,berger_Annu.Rev.FluidMech._1972}.
Moreover, when the drag force $F_\mathrm{D}$ is normalized to the coefficient  $C_{\mathrm{D}}=F_{\mathrm{D}}/(\tfrac{1}{2}\rho U^{2}A)$, where $\rho$ is the mass density of the fluid and $A$ is the reference area of the object,
it exhibits well--known dependence on the Reynolds number~\cite{oertel_Annu.Rev.FluidMech._1990}, 
typically taking values of $C_\mathrm{D}\sim1$ over a broad range of $\mathrm{Re}\sim10^3-10^5$ \cite{Lienhard_1966}.

The advent of quantum fluids has enabled tests of such universality of dynamic similarity in new regimes.  
Liquid helium systems have long served as a platform for research on inviscid flows~\cite{kapitza_Nature_1938,allen_Nature_1938,rayfield_Phys.Rev._1964}, 
and ultracold atomic gases now provide an especially clean and highly controllable system, 
where versatile flow geometries can be engineered by optical methods with high precision. 
In these systems, one can investigate the properties of dissipationless fluids and probe exotic phenomena unique to superfluids, such as the nucleation and dynamics of quantized vortices~\cite{landau_1987, ramanathan_Phys.Rev.Lett._2011,neely_Phys.Rev.Lett._2010,neely_Phys.Rev.Lett._2013,matthews_Phys.Rev.Lett._1999,raman_Phys.Rev.Lett._1999}.  
An increasing number of studies have highlighted parallels between classical and quantum flows.
The von K\'arm\'an vortex streets behind obstacles~\cite{sasaki_Phys.Rev.Lett._2010,stagg_2014,kwon_Phys.Rev.Lett._2016,kwak_Phys.Rev.A_2023,li_NewJ.Phys._2019}, the scaling law in quantum turbulence~\cite{navon_Nature_2016,johnstone_Science_2019} and universal scaling along with the superfluid Reynolds number in shedding dynamics~\cite{reeves_Phys.Rev.Lett._2015,christenhusz_Phys.Rev.Lett._2025,lim_NewJ.Phys._2022} represent cases in which well--known concepts in classical fluid dynamics can be extended to quantum cases. 
Recently, it was theoretically shown that dissipation in classical fluids can be understood through the Josephson--Anderson relation for superfluids~\cite{eyink_Phys.Rev.X_2021}, demonstrating that concepts pertaining to superfluid physics can provide valuable insight into classical fluid dynamics.

In this work, we investigate dynamic similarity in the wake dynamics in a superfluid, focusing on a regime involving a {\it penetrable} obstacle.
The ability of the fluid to penetrate the obstacle fundamentally modifies the vortex nucleation process~\cite{kwak_Phys.Rev.A_2023,kwon_Phys.Rev.A_2015}.
In contrast to impenetrable obstacles, which nucleate individual vortices from  the boundary of density-depleted regions, penetrable obstacles produce vortex dipoles through a precursor rarefaction pulse.
This penetrable obstacle has no true classical analogue, but
it is readily realized in atomic Bose-Einstein condensate (BEC) experiments using a focused laser beam whose potential strength experienced by the atoms is below the chemical potential of the BEC~\cite{kwak_Phys.Rev.A_2023,kwon_Phys.Rev.A_2015,lim_NewJ.Phys._2022, engels_Phys.Rev.Lett._2007}.
There have been experimental studies of penetrable obstacles in which the periodic generation of vortex dipoles from a moving obstacle was observed~\cite{lim_NewJ.Phys._2022, kwon_Phys.Rev.A_2015} and the quantitative relationship between the dipole generation rate and the obstacle speed and strength was investigated~\cite{lim_NewJ.Phys._2022}. 
Because the dynamic similarity in wake dynamics was numerically presented for impenetrable obstacles~\cite{reeves_Phys.Rev.Lett._2015}, 
it is therefore natural to expect that this concept can be extended to penetrable obstacles.
However, a unifying quantitative framework for vortex shedding across different regimes is still lacking,
particularly because it is difficult to define the length scale of the penetrable obstacle in an unambiguous manner.

Here, we analyze shedding dynamics from a penetrable obstacle by numerically solving the Gross--Pitaevskii equations (GPEs) and present an appropriate \textit{effective length scale} that captures the evolution of the shedding behaviors. 
We define the effective diameter from the Mach-1 contour of the time-averaged irrotational flow around the penetrable obstacle, which delineates the locally supersonic region where quantized vortices nucleate.
We show that, once this length is incorporated into a new superfluid Reynolds number, the resulting dimensionless quantities, in this case the Strouhal number and drag coefficient, exhibit universal behaviors in close analogy with classical fluids.  
Our findings thus extend the scope of dynamic similarity to penetrable superfluid flows, showing that the framework of dynamic similarity extends to this quantum regime.

\section{Method}
We consider a situation in which a BEC flows at constant velocity $\bm{v}_\infty = -v_0 \hat{\mathbf{x}}$ past an obstacle potential $V(\mathbf{r})$.
To simulate the superfluid vortex shedding induced by the obstacle, we numerically solve the two-dimensional GPE for the condensate wave function $\psi(\mathbf{r},t)$, which is given by 
\begin{equation}
i \hbar \frac{\partial \psi}{\partial t}=\left(-\frac{\hbar^2}{2 m} \nabla^2+i \hbar v_0 \frac{\partial}{\partial x}+V(\mathbf{r})+g|\psi|^2\right) \psi,
\label{eq:gpe}
\end{equation}
where $m$ is the mass of the particles and $g$ is the interaction strength.
The wave function can be expressed as 
$\psi(\mathbf{r},t)=\sqrt{\rho(\mathbf{r},t)}\,e^{i\theta(\mathbf{r},t)}$, where $\rho$ and $\theta$ denote the density and phase of the condensate, respectively, and the local superfluid velocity is expressed as $\bm v(\mathbf r,t) = (\hbar/m)\nabla \theta$. 
The obstacle is modeled by Gaussian potential
\(
V(\mathbf{r}) = V_0 \, \exp\left[-2(x^2+y^2)/\sigma^2\right],
\)
where \(V_0(>0)\) is the peak barrier height and
\(\sigma\) is the \(1/e^{2}\) radius.
For a penetrable obstacle, we take \(V_0<\mu\), where $\mu=g \rho_0$ is the chemical potential of the condensate, such that no fully depleted density core forms, in contrast to the impenetrable case~\cite{kwak_Phys.Rev.A_2023}.

The initial trial wavefunction was prepared from the Thomas--Fermi approximation in the presence of the obstacle and relaxed to the true ground state for a stationary BEC by imaginary-time propagation~\cite{antoine_ComputerPhysicsCommunications_2014}.
The real-time evolution is calculated using the Fourier pseudo-spectral method~\cite{antoine_ComputerPhysicsCommunications_2015} with the fourth-order Runge-kutta method with $\Delta t=0.01\tau$ and $\tau=\hbar/\mu$.
During the real-time evolution, the background flow was linearly increased to target velocity $v_0$ over a short acceleration interval of \(0.1\,\tau\). 
The simulation domain is $-350<x/\xi<150$ and $|y/\xi|<125$ and the spatial resolution is $0.5\xi$ per grid point, where $\xi=\hbar/\sqrt{m\mu}$ is the condensate healing length.

To track the long-time evolution until the wake pattern reaches a steady state, we utilize the fringe region technique~\cite{frisch_Phys.Rev.Lett._1992,reeves_Phys.Rev.Lett._2015}
in the edge region of the computational domain.
The fringe region serves to absorb the propagating density undulation, preventing interference from waves.
This is implemented with damping function $\gamma(\mathbf{r})$.
We set $\gamma(\mathbf r) = \max\!\big[\gamma(x), \gamma(y)\big]$, where
\begin{equation}
\gamma(\alpha) = \frac{\gamma_0}{2} \left\{ 2 + \tanh\!\left(\frac{\alpha-w_\alpha}{d}\right) - \tanh\!\left(\frac{\alpha+w_\alpha}{d}\right) \right\},
\label{eq:dampedGPE}
\end{equation}
with $\alpha=x,y$, $\gamma_0=0.1$, $w_x=25\xi$, $w_y=20\xi$ and $d=3\xi$.
Furthermore, to prevent vortex reentry under the periodic boundary conditions, we use the vortex-unwinding algorithm to remove the vortex adjacent to the boundary region~\cite{billam_Phys.Rev.Lett._2014,reeves_Phys.Rev.Lett._2015}. 
Vortex removal is implemented every $5\tau$ during time evolution.

To investigate the long-time steady-state shedding regime, we examined the vortex shedding over a window of \(2000 \tau\le t \le 10000\tau\), thus excluding the initial acceleration stage. 
To break symmetry during initial shedding, we added a small amount of complex random noise to the initial wave function after the imaginary-time evolution, with uniformly distributed real and imaginary parts scaled to a relative amplitude of the order of $10^{-4}$.

\section{Results}

\subsection{Vortex dipole--to--cluster transition} \label{subsec:A}
\begin{figure*}[t!]
    \centering
    \includegraphics[width=\linewidth]{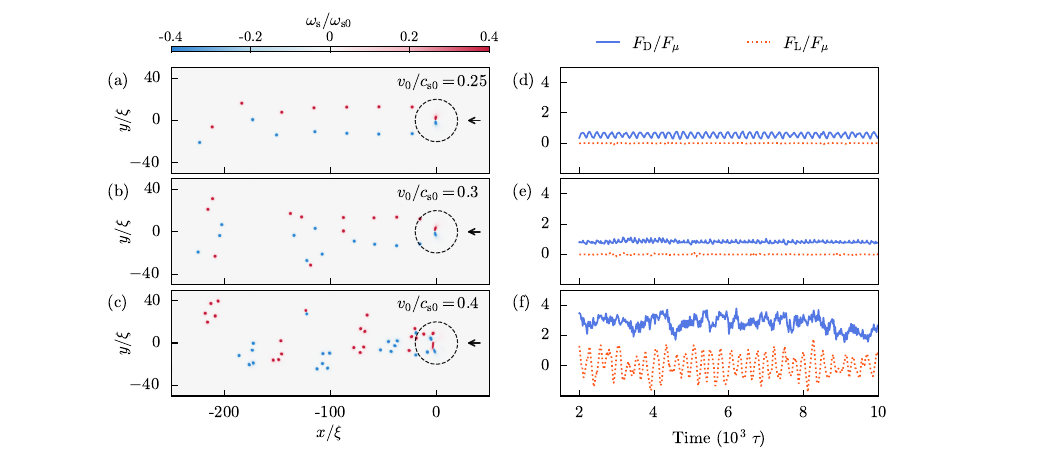}
    \caption{
Vortex shedding in a Bose-Einstein condensate flowing past a penetrable obstacle. The condensate flows from right to left (black arrow) at a speed of $v_0$. The Gaussian obstacle has strength $V_0/\mu = 0.9$ and width $\sigma/\xi = 20$.
(a)–(c) Superfluid vorticity distributions $\omega_\mathrm{s}(\mathbf{r})$ for $v_0/c_{s0}=0.25$ (a), $0.30$ (b), and $0.40$ (c), where $c_{s0}$ is the speed of sound in the unperturbed condensate.
$\omega_\mathrm{s}(\mathbf{r})$ is normalized by $\omega_\mathrm{s0}=\rho_0/\tau$.
As $v_0$ increases, shedding evolves from periodic vortex-dipole emission to same-sign vortex-cluster shedding.
Dashed circles indicate the obstacle's $1/e^2$ radius, centered at $x=0$. 
(d)–(f) Time traces of drag force $F_{\mathrm{D}}$ (solid blue) and lift force
$F_{\mathrm{L}}$ (dotted red) for the same parameters used in (a)–(c).
The force is normalized by $F_{\mathrm{\mu}}=\mu/\xi$.
As $v_0$ increases, $F_\mathrm{D}$ grows in magnitude and becomes increasingly irregular, while $F_\mathrm{L}$ increases in magnitude and develops stronger oscillations. Together, these trends reflect the transition from dipole- to cluster-dominated shedding.
} 
    \label{fig:wake_pattern}
\end{figure*}

\begin{figure}
    \centering
    \includegraphics[width=\linewidth]{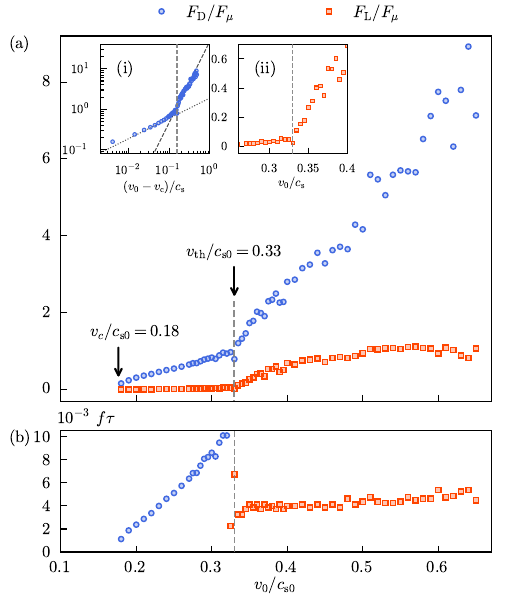}
    \caption{Drag and lift responses for a penetrable obstacle with
$V_0/\mu = 0.9$ and $\sigma/\xi = 20$.
(a) Mean drag $\overline{F_{\mathrm{D}}}$ (blue circles) and the root-mean-square of the lift $(\overline{F_\mathrm{L}^2})^{1/2}$ (red squares) versus background flow speed $v_0/c_{s0}$. 
The critical velocity $v_c=0.18 c_\mathrm{s0}$ and the dipole-to-cluster transition velocity $v_\text{th}= 0.33 c_\mathrm{s0}$ are indicated by the black arrow and the grey dashed line, respectively.
Left inset: log--log plot of $F_{\mathrm{D}}$ versus $(v_0 - v_c)$ with guide lines $(v_0-v_c)^{1/2}$ (dotted) and $(v_0-v_c)^2$ (dashed). 
Right inset: magnified view of $F_\mathrm{L}$ near the transition, showing a rapid rise for $v_0>v_\mathrm{th}$.
(b) Drag and lift frequencies $f_\mathrm{D}$ and $f_\mathrm{L}$ as functions of $v_0/c_{s0}$. 
The frequencies are normalized by $1/\tau = \mu/\hbar$.
Reliable determinations were obtained for $f_\mathrm{D}$ when $v_0 < v_{\mathrm{th}}$ and for $f_\mathrm{L}$ when $v_0 > v_{\mathrm{th}}$.}
    \label{fig:force_summary}
\end{figure}

To characterize the shedding behavior for a penetrable obstacle, we evaluated two quantities from numerical simulations: the spatial distribution of superfluid vorticity $\omega_\mathrm{s}(\mathbf{r})$ and the temporal evolution of the force $\mathbf{F}(t)$ exerted by the obstacle potential.
The superfluid vorticity is defined as the curl of the supercurrent density~\cite{ronning_Phys.Rev.Research_2023},
\begin{equation}
\omega_\mathrm{s}=\frac{1}{2}\nabla\times(\rho\mathbf{v})
  =\frac{1}{2i}\!\left(\partial_x\psi^*\,\partial_y\psi
  -\partial_y\psi^*\,\partial_x\psi\right),
\end{equation}
providing a convenient measure of vortex generation, annihilation, and propagation around the obstacle.
The force exerted by the obstacle is estimated as
\begin{equation}
\mathbf{F}(t)
= -\iint_\Omega dx\,dy\,|\psi(\mathbf{r},t)|^2
   \nabla\!\left[\frac{V(\mathbf{r},t)}{m}\right], 
\label{eq:force}
\end{equation}
where $\Omega$ is the two–dimensional computational domain. The force is decomposed as $\mathbf{F}(t)=F_\mathrm{D}(t)\hat{\mathbf{x}} + F_\mathrm{L}(t)\hat{\mathbf{y}}$, where $F_\mathrm{D}$ and $F_\mathrm{L}$ are the drag and lift components, parallel and perpendicular to the background flow, respectively.

In Fig.~\figref{fig:wake_pattern}, we display $\omega_\mathrm{s}(\mathbf{r})$ and $\mathbf{F}(t)$ for different flow velocities with $V_0/\mu=0.9$ and $\sigma/\xi=20$. 
Just above the critical velocity $v_\mathrm{c}\approx 0.18 c_{\rm s0}$, where $c_{\rm s0}=\sqrt{\mu/m}$ is the speed of sound in the unperturbed condensate, vortex dipoles are nucleated within the penetrable obstacle via phase accumulation and phase slip [Fig.~\figref[a]{fig:wake_pattern}]~\cite{kwak_Phys.Rev.A_2023}.  
These dipoles form nearly parallel rows as they propagate downstream.  
Because such dipole rows are unstable~\cite{Lamb_1945, sasaki_Phys.Rev.Lett._2010}, the downstream wake gradually reorganizes into a staggered V–shaped pattern~\cite{li_NewJ.Phys._2019}.  
As the velocity increases, this instability spreads upstream toward the obstacle, resulting in increasingly irregular vortex configurations in the near wake [Fig.~\figref[b]{fig:wake_pattern}].  
At still higher velocities, the nucleation mechanism remains predominantly pair production at the center of the obstacle, while the increased rate of pair-production raises the vortex density in the wake.
The resulting high vortex density promotes vortex-vortex interactions, leading to annihilation events and the formation of Jones--Roberts solitons~\cite{jones_J.Phys.A:Math.Gen._1982} [Fig.~\figref[c]{fig:wake_pattern}].  
These interactions lead to the formation of same–sign vortex clusters, which persist downstream and may eventually organize into a quasi–von Kármán street~\cite{yoneda_J.Phys.Soc.Jpn._2025}, despite the fact that the obstacle is penetrable (see Appendix A).

The corresponding force response, shown in Figs.~\figref[d]{fig:wake_pattern}--\figref[f]{fig:wake_pattern}, reflects the same progression.  
In the dipole–row regime near $v_\mathrm{c}$, the drag force exhibits periodic oscillations with negligible lift, consistent with symmetric dipole nucleation.  
As the velocity increases, the drag-oscillation frequency rises, and the drag force shows larger fluctuations, in line with the growing instability in the wake.  
Once vortex clusters appear, the drag force becomes strongly irregular, while the lift force develops pronounced periodic oscillations associated with alternating cluster formation [Fig.~\figref[f]{fig:wake_pattern}].
These results show that, as the flow velocity increases, the shedding behavior of a penetrable obstacle changes from periodic dipole–row emission to the shedding of same–sign vortex clusters. These observations are consistent with earlier numerical studies~\cite{Reeves_PRA_2012}.

We investigate the dipole–to–cluster transition further by examining the velocity dependence of the drag and lift forces. Figure~\figref[a]{fig:force_summary} shows the time-averaged drag force $\overline{F_\mathrm{D}}$ and the root-mean-square of the lift force, $(\overline{F_\mathrm{L}^2})^{1/2}$ as functions of $v_0$. 
Here, the time averages are evaluated for $2000\le t/\tau\le10000$ with snapshots
sampled every \(\Delta t = 10\,\tau\). 
Both the drag and lift exhibit sudden increases at the threshold velocity, $v_\mathrm{th}\simeq0.33\,c_{\mathrm{s0}}$, which is also associated with the onset of asymmetric vortex cluster shedding. 
We find that the drag force closely follows the scaling $\overline{F_\mathrm{D}}\sim (v_0-v_\mathrm{c})^{1/2}$ in the low-velocity dipole-emission regime~\cite{huepe_PhysicaD:NonlinearPhenomena_2000}, while it crossed over to a quadratic dependence $\overline{F_\mathrm{D}}\sim (v_0-v_\mathrm{c})^{2}$ in the high-velocity cluster-shedding regime [Fig.~\figref[a]{fig:force_summary} inset].

The dipole-to-cluster transition is also examined by analyzing the oscillatory behavior of the drag and lift forces
through their Fourier spectra. 
The drag force displays a well-defined shedding frequency up to $v_\mathrm{th}$, consistent with the periodic dipole emission, while above $v_\mathrm{th}$, the spectrum becomes irregular and no dominant frequency remains. In contrast, the lift force shows a negligible spectral weight below $v_\mathrm{th}$ but develops a clear oscillation frequency above $v_\mathrm{th}$, reflecting the alternating release of vortex clusters. 
From the spectral peaks of the Fourier spectra of $F_\mathrm{D}(t)$ and $F_\mathrm{L}(t)$, we determine their characteristic oscillating frequencies $f_\mathrm{D}$ and $f_\mathrm{L}$, respectively. 
The measurement results are shown in Fig.~\figref[b]{fig:force_summary} for the corresponding shedding regimes across the threshold velocity. 
In the dipole shedding regime, $f_\mathrm{D}$ increases almost linearly with an increase of $v_0$ above $v_c$, consistent with previous experiments~\cite{lim_NewJ.Phys._2022, kwon_Phys.Rev.A_2015}. By contrast, in the cluster shedding regime, $f_\mathrm{L}$ exhibits very weak dependence on $v_0$. 
Near the threshold velocity $v_\mathrm{th}$, the measured frequencies suggest an approximate relation $f_\mathrm{L}\sim f_\mathrm{D}/2$, which is consistent with the onset of alternating, symmetry-broken shedding.

\subsection{Effective Diameter}\label{sec:B}
\begin{figure}
    \centering
\includegraphics[width=\linewidth]{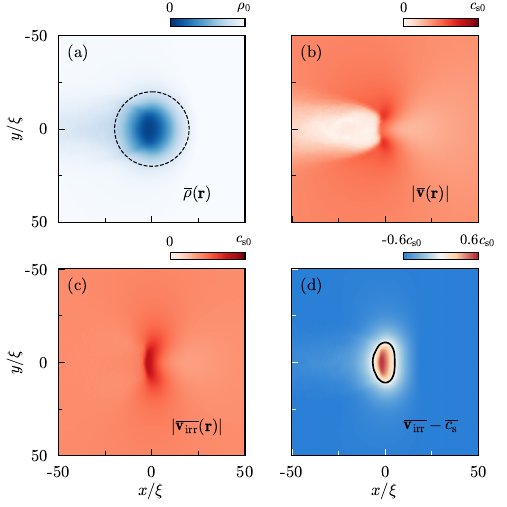}
    \caption{%
    Time-averaged flow fields around a penetrable obstacle for $V_0/\mu=0.9$, $\sigma/\xi=20$, and $v_0/c_{\mathrm{s0}}=0.4$.
    (a) Condensate density $\overline{\rho}(\mathbf{r})$; the dashed circle indicates the obstacle's $1/e^2$ radius. 
    (b) Flow speed $|\overline{\bm{v}}(\mathbf{r})|$ and (c) irrotational flow speed $|\overline{\bm{v}_\mathrm{irr}}(\mathbf{r})|$.
    (d) Difference $|\overline{\bm{v}_\mathrm{irr}}(\mathbf{r})|- \overline{c_{\mathrm{s}}}(\mathbf{r})$, where $\overline{c_{\mathrm{s}}}(\mathbf{r})$ is the local speed of sound as determined by $\overline{\rho}(\mathbf{r})$.
    The black contour is the Mach-1 contour   
    $|\overline{\bm{v}_\mathrm{irr}}(\mathbf{r})|=\overline{c_{\mathrm{s}}}(\mathbf{r})$.
}
    \label{fig:supersonic}
\end{figure}

\begin{figure*}[t!]
\centering
\includegraphics[width=\linewidth]{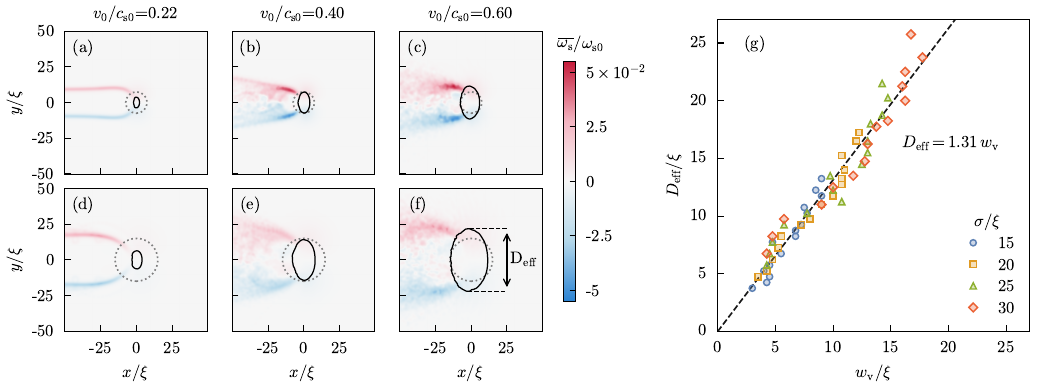}
\caption{%
    Time-averaged vorticity fields $\overline{\omega_\mathrm{s}}(\mathbf{r})$ for various flow speeds $v_0/c_{\rm s0}$; (a-c) $V_0/\mu=0.9$ and $\sigma/\xi=20$, and (d-f) $V_0/\mu=0.9$ and $\sigma/\xi=40$. 
    The black solid line denotes the Mach-1 contour $\overline{v_{\rm irr}}(\mathbf r)=\overline{c_{\rm s}}(\mathbf r)$, enclosing the locally supersonic region, and the gray dotted circle marks the obstacle's $1/e^2$ radius.
    The effective diameter $D_{\mathrm{eff}}$ is defined as the maximum extent of the Mach-1 region along the $y$ direction.
    (g) Effective diameter $D_{\rm eff}/\xi$ versus vorticity width $w_{\rm v}/\xi$. The width $w_{\rm v}$ was determined from $\overline{\omega_\mathrm{s}}(\mathbf{r})$ (see text). Data points are shown for various values of $\sigma$ and $0.22\leq v_0/c_{\rm s0} \leq 0.65$ at \(V_0/\mu=0.9\). Error bars are smaller than the symbols. The dashed line is a linear fit to the data. }
\label{fig:effective_diameter}
\end{figure*}

Our analysis of the wake morphology and force response clearly shows a distinct dipole–to–cluster transition in the superfluid shedding dynamics for a penetrable obstacle.
In this section, we introduce the \textit{effective diameter} $D_{\rm eff}$ for a penetrable obstacle to determine the characteristic length scale relevant to the shedding dynamics, enabling the definition of a proper superfluid Reynolds number $\mathrm{Re}_s$ to organize the shedding behavior within the framework of dynamic similarity.

Vortex nucleation in a superfluid occurs when the local flow speed reaches or exceeds the local speed of sound~\cite{kokubo_pra_2025, josserand_PhysicaD:NonlinearPhenomena_1999, musser_Phys.Rev.Lett._2019, rica_2001, winiecki_2000}. 
This condition is referred to as the local Landau criterion and is expressed as
\begin{equation}
    v(\mathbf{r}) > c_\mathrm{s}(\mathbf{r}) = \sqrt{g \rho(\mathbf{r})/m} ,
    \label{eq:compressibility1}
\end{equation}
where $c_{\rm s}(\mathbf{r})$ denotes the local speed of sound. This suggests that the spatial region around the obstacle where the flow becomes locally supersonic can be considered as the effective spatial extent of the obstacle in terms of vortex shedding. 
We thus define the effective obstacle diameter $D_{\rm eff}$ as the transverse width of the locally supersonic region (along the $y$--direction).

To gain insight into the flow structure around a penetrable obstacle, we examine the spatial distributions averaged over time of the density $\overline{\rho}(\mathbf{r})$ and velocity $\overline{\bm{v}} (\mathbf{r})$. The exemplary data in the case of $v_0>v_\mathrm{th}$ are shown in Fig.~\figref{fig:supersonic}.
The density of condensate $\overline{\rho}$ exhibits a slight suppression compared to that of a stationary flow behind the obstacle due to the trail of the shed vortices downstream [Fig.~\figref[a]{fig:supersonic}].
The average flow speed $|\overline{\bm{v}}|$ in Fig.~\figref[b]{fig:supersonic} clearly reflects the influence of these vortices: downstream of the obstacle, a low--velocity region develops where the background flow is partially canceled by the vortex-induced flow, while on the upstream side of the obstacle center the flow velocity is locally enhanced.

Given that we are interested in determining the supersonic region {\it before} vortex shedding, it is desirable to estimate the velocity field without the chaotic wake induced by shed vortices. This can be done by calculating the compressible (irrotational) component of the velocity field \(\bm v_{\rm irr}(\mathbf r)\)~\cite{navon_Nature_2016, reeves_Phys.Rev.Lett._2015, musser_Phys.Rev.Lett._2019}.
According to Helmholtz decomposition, $\bm{v} = \bm{v}_{\mathrm{irr}} + \bm{v}_{\mathrm{sol}}$ with $\nabla \times \bm{v}_{\mathrm{irr}} = 0$ and $\nabla \cdot \bm{v}_{\mathrm{sol}} = 0$.Since the vortex contribution is confined to the solenoidal component \(\bm{v}_{\mathrm{sol}}\), \(\bm v_{\rm irr}\) provides a clear separation between compressibility-driven instabilities and vortex-induced fluctuations. As vortex nucleation is initiated by instabilities associated with compressibility, it is therefore more physically meaningful to define \(D_{\rm eff}\) in terms of the compressible component.

In Fig.~\figref[c]{fig:supersonic}, we observe that the time–averaged irrotational flow speed reaches its maximum at the obstacle center.
This means that phase slips caused by vortex nucleation, which relax the accumulated phase gradient, lead to an overall reduction of $|\overline{\bm{v}(\rm r)}|$ as shown in Fig.~\figref[b]{fig:supersonic}.
Consequently, we define the supersonic region as 
\begin{equation}
\mathcal S \equiv \bigl\{\mathbf r\in\Omega:\;
\lvert\overline{\bm v_{\rm irr}}(\mathbf r)\rvert \ge
\overline{c_s}(\mathbf r)\bigr\},
\label{eq:D_eff}
\end{equation}
with $\overline{c_s}(\mathbf r)=\sqrt{g \overline{\rho}(\mathbf r)/m }$ and determine \(D_{\rm eff}\) to be the transverse extent of \(\mathcal S\) through the obstacle center.
In other words, $D_\mathrm{eff}$ corresponds to the transverse size of the Mach--1 contour around the obstacle [black solid line in Fig.~\figref[d]{fig:supersonic}].
We note that there is an alternative way to determine an effective diameter $D_{\rm eff}$ from the supersonic region $\mathcal{S}$. For example, the size of $\mathcal{S}$ can be characterized in terms of its area $A_\mathcal{S}$, i.e., the equivalent diameter $2\sqrt{A_\mathcal{S}/\pi}$.
A systematic comparison of such alternatives is left for future work.

To test whether the supersonic region identified through \(\overline{\bm v_{\rm irr}}\) suitably represents the vortex nucleation region, we compare it with the spatial distribution of the vortices.  
Figures~\figref[a]{fig:effective_diameter}--\figref[f]{fig:effective_diameter} show the vorticity field averaged over time, \(\overline{\omega_\mathrm{s}}(\mathbf{r})\), together with the supersonic region defined by Eq.~\eqref{eq:D_eff}.
Trails of positive (blue) and negative (red) vorticity appear just downstream of the obstacle. 
As the velocity increases, the vorticity-concentrated region expands and the vortex trail behind the obstacle grows into an enlarged wake along with the expansion of the supersonic region. 
$\overline{\omega_\mathrm{s}}$ is predominantly concentrated along the lateral boundaries of the supersonic zone~\cite{footnote_streetwidth}, in contrast to the fixed Gaussian width of the obstacle.

In Fig.~\figref[g]{fig:effective_diameter}, we plot the effective diameter $D_{\rm eff}$ for various Gaussian widths $\sigma$ and flow speed $v_0$ compared to the lateral width $w_\mathrm{v}$ of the distribution $\overline{\omega_\mathrm{s}}(\mathbf{r})$.
To obtain $w_\mathrm{v}$, we integrate $\overline{\omega_\mathrm{s}}(\mathbf{r})$ along the $x$--direction within a finite region that encompasses the supersonic zone and extends downstream by $20\xi$.
The width $w_\mathrm{v}$ is then defined as the distance between the positive and negative peaks in the projected $\omega_\mathrm{s}$ profile.
The effective diameter shows a linear relationship with $w_\mathrm{v}$, $D_{\rm eff}\approx 1.31 w_\mathrm{v}$, for a wide range of obstacle conditions, showing that the supersonic region sets the length scale of the penetrable obstacle for vortex shedding. 
An analytical description of the $v_0$-dependence of $D_\mathrm{eff}$ is presented in Appendix B.

From the observation of $D_{\rm eff} \propto w_\mathrm{v}$ and motivated by Reeves et al.~\cite{reeves_Phys.Rev.Lett._2015}, we propose a superfluid Reynolds number for penetrable obstacles, as follows:
\begin{equation}
    \mathrm{Re}_{\rm s} = \frac{(v_0 - v_{\rm c}) D_{\rm eff}}{\hbar/m}.
    \label{eq:threshold_Re}
\end{equation}
Physically, $\mathrm{Re_s}$ can be interpreted as the circulation of the background flow around the obstacle, measured in units of the circulation quantum $\kappa = h/m$, up to a numerical prefactor. 
This provides an estimate of the number of vortices that the flow can sustain around the obstacle.
The replacement of $v_0$ by $v_0 - v_c$ reflects the fact that dissipation is strictly absent below $v_c$, implicitly encoding obstacle-specific geometric details into the similarity parameter~\cite{reeves_Phys.Rev.Lett._2015}.
In computing $\textrm{Re}_{\rm s}$, we determined the critical velocity \(v_c\), following Refs.~\cite{kwak_Phys.Rev.A_2023,huynh_Phys.Rev.A_2024}, by using the imaginary-time propagation method to check the existence
of a stationary ground-state solution for $v_0<v_c$.

\subsection{Superfluid Reynolds number $\text{\rm Re}_{\rm s}$ and Strouhal number $\text{\rm St}$}
\begin{figure}
    \centering
    \includegraphics[width=\linewidth]{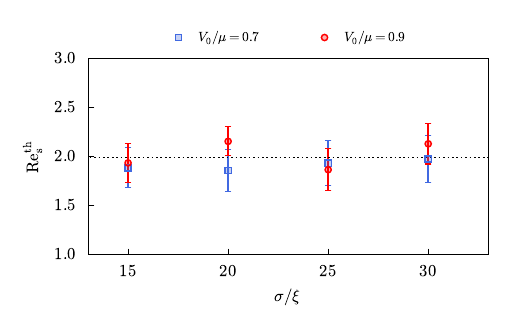}
    \caption{Dynamic similarity of the dipole-to-cluster transition.
Threshold superfluid Reynolds number $\mathrm{Re}_{\rm s}^{\rm th}$ at $v_0=v_\textrm{th}$ for obstacle parameters $V_0/\mu \in \{0.7, 0.9\}$ and $\sigma/\xi \in \{15, 20, 25, 30\}$. 
Error bars indicate the uncertainties of $ \mathrm{Re}_s^{\mathrm{th}} $, arising from the finite numerical resolution in determining $v_c$, $v_{\rm th}$ and $D_\textrm{eff}$. 
The black dotted line indicates the mean value of all data points.
}
    \label{fig:placeholder}
\end{figure}

First, we examine whether the superfluid Reynolds number $\mathrm{Re}_{\rm s}$ can serve as a reliable indicator of the dipole--to-cluster transition. 
In Fig.~\figref{fig:placeholder}, we plot the values of  $\mathrm{Re}_{\rm s}^{\rm th}$ at $v_0=v_\textrm{th}$ for various obstacle conditions in terms of $\sigma$ and $V_0/\mu$.
Here, $v_{\rm th}$ is determined as the velocity above which the drag force spectrum no longer shows a peak structure, which is fully correlated with the onset behavior of the lift force.
It is remarkable that the dipole-to-cluster transition occurs consistently at $\mathrm{Re}_{\rm s}\approx 2$. 
This shows that the superfluid Reynolds number defined in Eq.~\eqref{eq:threshold_Re} conveys the dynamic similarity in the wake dynamics for penetrable obstacles.
We note that as the Gaussian width $\sigma$ increases from 15$\xi$ to 30$\xi$, the ratio of $D_{\rm eff}/\sigma$ decreases by a factor of 1.6, showing that $D_{\rm eff}$ appropriately represents the length scale of the obstacle.

Next, we examine the relationship between $\mathrm{Re}_s$ and the Strouhal number $\mathrm{St}$ for penetrable obstacles. Using the effective diameter, we compute $\mathrm{St}$ as follows:
\begin{equation}
\mathrm{St}\;=\; \frac{f_s\,D_{\mathrm{eff}}}{\,v_0}.
\label{eq:St_def}
\end{equation}
Here, the vortex shedding frequency \(f_s\) is determined via
\begin{equation}
f_s(v_0) \;=\;
\begin{cases}
\displaystyle f_{\!L}, & v_0 \ge v_{\rm th},
\\[6pt]
\displaystyle \tfrac{1}{2}\,f_{\!D}, & v_0 < v_{\rm th}.
\end{cases}
\label{eq:fs_piecewise}
\end{equation}
For $v_0 < v_{\rm th}$, the lift spectrum lacks a clear dominant peak; thus, we approximate $f_s$ as half the drag frequency. This approach is motivated by the classical low-\(\mathrm{Re}\) result showing that the lift frequency is half the drag frequency~\cite{norberg_JournalofFluidsandStructures_2003}, and is supported by earlier studies of impenetrable obstacles~\cite{sasaki_Phys.Rev.Lett._2010,li_NewJ.Phys._2019}.

\begin{figure}
    \centering
    \includegraphics[width=\linewidth]{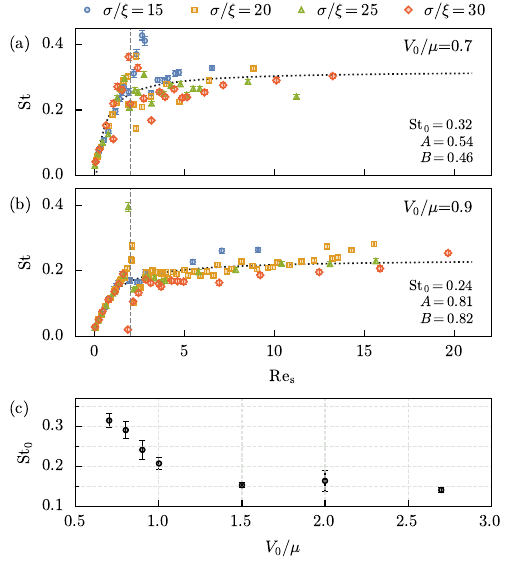}
    \caption{
    Strouhal number $\mathrm{St}$ versus $\mathrm{Re}_{\rm s}$ for $V_0/\mu=0.7$ (a) and $0.9$ (b). $\mathrm{St}$ was computed using $D_\text{eff}$ [Eq.~\eqref{eq:St_def}]. 
    Vertical dashed lines indicate $\mathrm{Re}_{\mathrm{s}} = 2$.
    Dotted curves are fits to $\mathrm{St}(\mathrm{Re}_{\rm s}) = \mathrm{St}_0 \!\left[ 1 - \frac{A}{\mathrm{Re}_{\rm s} + B} \right]$; fit parameters are listed in each panel.
    Error bars represent the uncertainty in $\mathrm{St}$ from the finite numerical resolution in determining $D_{\mathrm{eff}}$ 
    and the FFT frequency resolution to extract $f_\mathrm{s}$.
    (c) Asymptotic Strouhal number $\mathrm{St}_0$ as a function of $V_0/\mu$. 
    $\mathrm{St}_0$ was obtained by averaging $\mathrm{St}$ over $0.55\leq v_0/c_\mathrm{s0} \leq 0.65$ where $\mathrm{St}$ is saturated ($\sigma/\xi=20$), and its error bar indicates the standard error of the mean.}
    \label{fig:Re-St}
\end{figure}

Figures~\figref[a]{fig:Re-St} and \figref[b]{fig:Re-St} present our numerical results for $V_0/\mu=0.7$ and 0.9, respectively, on the plane of $\mathrm{Re}_s$ and $\mathrm{St}$.
For a fixed obstacle strength $V_0/\mu$, data points of various obstacle sizes collapse onto a single curve.
Moreover, we observe that $\mathrm{St}$ remains nearly constant for $\mathrm{Re_s}>2$, representing a close analogy to the classical case where $\mathrm{St}$ saturates in the turbulent vortex shedding regime \cite{Lienhard_1966}.
The $\mathrm{Re}_{\rm s}$--$\rm St$ trend is well described by the empirical fit~\cite{roshko_1954, reeves_Phys.Rev.Lett._2015}
\begin{equation}
    \mathrm{St}(\mathrm{Re}_{\rm s}) = \mathrm{St}_0 \!\left[ 1 - \frac{A}{\mathrm{Re}_{\rm s} + B} \right],
    \label{eq:St_fit}
\end{equation}
which closely resembles the classical formulation.  
Within the limit of $\mathrm{Re_s} \gg 1$, $\mathrm{St}$ asymptotically approaches $\mathrm{St_0} \simeq 0.20$–$0.30$, remarkably close to the classical value of $\mathrm{St_0\simeq0.21}$~\cite{roshko_1954,Lienhard_1966}.

To understand how $\mathrm{St_0}$ depends on the obstacle strength, we extend our analysis across a wide range of barrier heights, including the impenetrable regime where $D_\textrm{eff}$ remains well-defined. As shown in Fig.~\figref[c]{fig:Re-St},
$\mathrm{St_0}$ gradually decreases from about $0.30$ at $V_0/\mu=0.7$ to around $0.15$ at $V_0/\mu=1.5$. Beyond this point, deeper in the impenetrable regime, $\mathrm{St_0}$ remains nearly constant at $\sim 0.15$, consistent with previous findings where the geometric obstacle diameter was used~\cite{reeves_Phys.Rev.Lett._2015}.
The smooth evolution of the asymptotic Strouhal number with $V_0/\mu$ suggests that the use of $\mathrm{Re}_{\rm s}$ provides a unified framework to characterize the penetrable-to-impenetrable crossover~\cite{kwak_Phys.Rev.A_2023}, which warrants future investigation.

The systematic increase of $\mathrm{St_0}$ with a decrease of $V_0/\mu$ in the penetrable regime likely reflects the decreasing robustness of vortex clustering for weaker obstacles. 
Because the value of $\mathrm{St}$ characterizes the geometric spacing of vortex clusters in the wake, 
larger values of $\mathrm{St}$ correspond to shorter inter-cluster distances ($\approx v_0/(2f_s)$). 
This leads to stronger overlaps between successive vortices with a finite core size, suppressing the formation of a stable vortex street.
In our simulations, a well-defined shedding frequency ceases to appear for $V_0/\mu<0.6$, consistent with previous observations~\cite{Reeves_PRA_2012}.
For sufficiently weak obstacles, the flow disturbance is too small to form regular shedding and the critical velocity approaches $c_{s0}$~\cite{kwon_Phys.Rev.A_2015, kwak_Phys.Rev.A_2023}.

\subsection{Drag force}\label{subsection:D}

\begin{figure}
    \centering
    \includegraphics[width = \linewidth]{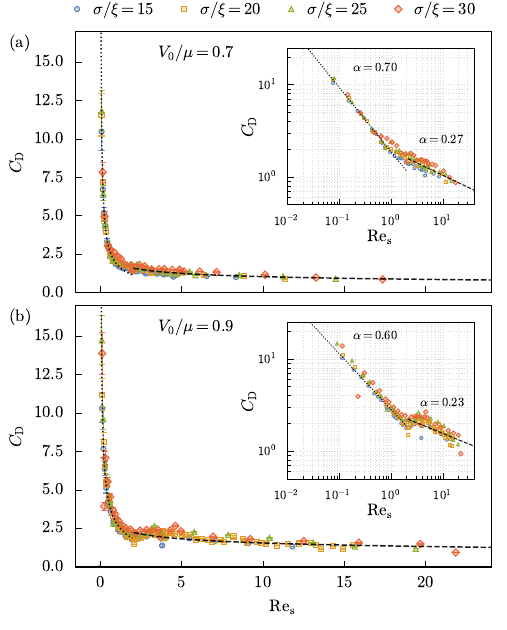}
    \caption{
    Drag coefficient $C_{\rm D}$ for a penetrable obstacle as a function of $\mathrm{Re_s}$. 
    $C_{\rm D}$ was estimated via Eq.~\eqref{eq:drag_fit}. 
    (a,b) Numerical data for $V_0/\mu=0.7$ (a) and $0.9$ (b).
    The dashed (dotted) curve is a fit to $C_{\mathrm{D}}=b\, \mathrm{Re_s}^{-\alpha}$ for $\mathrm{Re}_\mathrm{s} \geq 2$ ($\mathrm{Re}_\mathrm{s} \leq 2$).
    Error bars represent the uncertainties of $C_\mathrm{D}$, arising from the finite numerical resolution in the determination of $v_c$ and $D_\textrm{eff}$.
    The insets display the same data on log--log axes, with fitted power-law exponents $\alpha$.
    }
    \label{fig:drag_coeff}
\end{figure}

Finally, we investigate the dynamic similarity of the drag force for a penetrable obstacle.
Following Ref.~\cite{christenhusz_Phys.Rev.Lett._2025} and replacing the geometric diameter \(D\) with the effective diameter \(D_{\rm eff}\), we introduce the superfluid drag coefficient for a penetrable obstacle as
\begin{equation}
    C_{\mathrm{D}}
    \;\equiv\;
    \frac{2\,\overline{F_\mathrm{D}}}{\rho_0\,\bigl(v_0^2 - v_{\rm c}^2\bigr)\,D_{\rm eff}} \, .
\label{eq:drag_fit}
\end{equation}
We note that this particular choice of drag coefficient is not unique in the context of a superfluid \cite{takeuchi_annalenderphysik_2025}.

In Fig.~\figref{fig:drag_coeff}, we display the measured values of $C_{\mathrm{D}}$ as a function of $\mathrm{Re}_{\rm s}$.  
For a fixed obstacle strength $V_0/\mu$, data obtained for different values of $\sigma/\xi$ collapse onto a single curve, demonstrating dynamic similarity.
At $\mathrm{Re_s}\approx 2$, the drag coefficient curve exhibits characteristic bending that is nearly independent of $\sigma/\xi$. This feature is associated with the dipole--to--cluster transition.

To quantify the dependence of $C_{\mathrm{D}}$ on $\mathrm{Re_s}$, we fit a power-law function of $C_{\mathrm{D}}=b\, \mathrm{Re_s}^{-\alpha}$ to the data for the two different ranges, $\mathrm{Re_s<2}$ and $\mathrm{Re_s}\geq 2$, separately [Fig.~\figref{fig:drag_coeff} insets].
For the low-$\rm Re_s$ dipole shedding regime, the exponent obtained is in the range of $\alpha=0.6\sim0.7$, similar to Stokes' law in the 2D cylinder case in classical fluid dynamics~\cite{footnote_drag_coeff, lamb_1911}.
For the high-$\mathrm{Re}$ cluster shedding regime, the exponent decreases to $\alpha=0.22\sim0.27$. Although $C_\mathrm{D}$ gradually decreases with an increase of $\mathrm{Re_s}$ in our parameter range, it is not known whether it vanishes as $\mathrm{Re_s}\rightarrow\infty$.
In classical fluids, it has been argued that the dissipation rate remains finite as $\mathrm{Re\rightarrow\infty}$, implying a nonvanishing drag coefficient.
However, studies in smooth boundary or periodic boundary conditions have shown that the dimensionless dissipation can decrease with an increase of $\mathrm{Re}$ \cite{cadot_PRE_1997,iyer_Phys.Rev.Lett._2025}, suggesting that the drag coefficient need not approach a finite constant at large $\mathrm{Re}$. 
How this behavior carries over to superfluids without intrinsic viscosity remains an interesting question.

\section{CONCLUSION AND OUTLOOK}

We have shown that vortex shedding past a penetrable obstacle in a two-dimensional superfluid can be organized by a flow-defined length scale, the effective diameter $D_{\mathrm{eff}}$.
Defined as the transverse extent of the region where the compressible component of the superflow exceeds the local speed of sound, $D_{\mathrm{eff}}$ delineates the Mach-1 contour that marks the onset of compressibility-driven instability.
$D_{\mathrm{eff}}$ expands with $v_0$ and quantitatively matches the observed width of the vorticity distribution, confirming that the dynamically relevant scale is that felt by the flow rather than the geometric obstacle size.
When forces and frequencies are normalized by $D_{\mathrm{eff}}$, the Strouhal and drag coefficients collapse onto universal curves versus the superfluid Reynolds number 
$\mathrm{Re}_s$, with a clear transition at $\mathrm{Re}_s \approx 2$.
Thus, $D_{\mathrm{eff}}$ provides a unified similarity length governing both the shedding dynamics and mean dissipation in penetrable obstacle flows. 

To test the broader applicability of our approach, we have also examined representative cases of impenetrable obstacles and non-Gaussian penetrable obstacles (Appendix D) and found scaling behavior similar to that for penetrable Gaussian obstacles. 
These results suggest that, 
provided the obstacle generates a bounded supersonic region and sustains vortex shedding, the effective-diameter framework remains useful over the parameter ranges examined here.
The observed collapses further indicate that the Mach-1 contour provides a practical and physically motivated definition of the relevant obstacle length scale.

Nevertheless, it is important to note that the framework has an intrinsic limitation: it breaks down for $v_0/c_\mathrm{s0}>1$, where the flow becomes globally supersonic and no bounded Mach-1 region can be identified. In that regime, dissipation is expected to be governed by shock dynamics rather than vortex shedding~\cite{engels_Phys.Rev.Lett._2007,dutton_Science_2001,kamchatnov_Phys.Rev.Lett._2008}, and a shock-based length scale may be relevant. 
In classical compressible flows, a shock layer controls the pressure drag past a bluff body \cite{liepmann_1957}.

The analogy between penetrable obstacles in a superfluid and classical porous bodies is worth noting. 
Porous obstacles permit partial through-flow, and their wake dynamics have been studied extensively~\cite{Castro_J.Fluid Mech._1971, Ledda_Phys.Rev.Fluids_2018, Lombardi_Phys.Rev.Fluids_2023}. 
In classical flows, the through-flow weakens the recirculation zone where vorticity accumulates and pushes it downstream, eventually suppressing vortex shedding. 
In our superfluid system, the locally supersonic region and the near wake play a qualitatively analogous role, and increasing penetrability similarly diminishes vortex shedding by allowing greater flow penetration.
However, the underlying mechanisms are distinct: vorticity in porous-body wakes is generated in viscous shear layers, whereas in superfluids it arises from the breakdown of superfluidity within the locally supersonic region.

Finally, we note that the universality revealed here can be tested experimentally. 
The shedding frequency or $\mathrm{St}$ can be measured directly using a conventional time-of-flight imaging, as shown in Refs.~\cite{kwon_Phys.Rev.Lett._2016,lim_NewJ.Phys._2022,kwon_Phys.Rev.A_2015}. 
In addition, the drag force can be measured without direct mechanical probes using a control-volume analysis, as suggested in Ref.~\cite{christenhusz_Phys.Rev.Lett._2025}.
Moreover, a recent proposal based on matter--wave optics~\cite{murthy_2019} suggests that both the superfluid velocity fields and the phase of the order parameter can be measured using phase--imprinting techniques~\cite{burger_Phys.Rev.Lett._1999,delpace_Phys.Rev.X_2022,denschlag_Science_2000}, making it feasible to visualize the Mach-1 contour and measure $D_\mathrm{eff}$ directly.
Such experimental methods would allow testing the predicted universal behaviors.

\section{acknowledgements}
We thank Haneul Kwak for sharing the simulation code, Myeonghyeon Kim for computing support, and J.~Park, D.~Bae, J.~Lee, S.~Kim, and J.~Jung for insightful discussions. 
This work was supported by the National Research Foundation of Korea (Grants No. RS-2023-NR077280, No. RS-2023-NR119928, and No. RS-2024-00413957).

\section{Data availavility}
The data that support the findings of this article are openly available~\cite{kwon_vortex_shedding_repo_2026}.

\section*{Appendix A: Details of dipole--to--cluster transition}
\begin{figure}
\centering
\includegraphics[width=\linewidth]{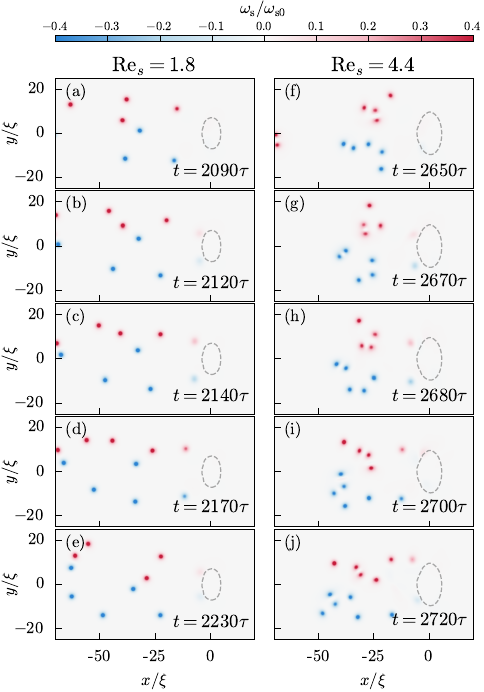}
\caption{%
Sequential snapshots of vortex nucleation near the obstacle for cases below and above the threshold $\mathrm{Re}_\mathrm{s}\approx 2$. 
(a--e) $\mathrm{Re}_\mathrm{s}=1.8$ $(v_0/c_{\mathrm{s}0}=0.31)$ and (f--j) $\mathrm{Re}_\mathrm{s}=4.4$ $(v_0/c_{\mathrm{s}0}=0.4)$, with $V_0/\mu=0.9$ and $\sigma/\xi=20$.
Each column shows successive stages of wake development: before vortex nucleation [(a),(f)],  nucleation event [(b),(g)], separation of opposite-sign vortices [(c),(h)], near-wake evolution [(d),(i)], and the start of the next cycle [(e),(j)].
The gray dashed line marks the Mach-1 contour.
}
\label{fig:pair_production}
\end{figure}

In Sec.~III A, we observed a sharp transition in the lift force near $\mathrm{Re}_\mathrm{s}\approx2$, coincident with the emergence of same-sign vortex clusters.
Figure~\figref{fig:pair_production} shows that vortices are nucleated by symmetric dipole pair production both below $\mathrm{Re}_\mathrm{s}<2$ [Figs.~\figref[a]{fig:pair_production}--\figref[e]{fig:pair_production}] and above $\mathrm{Re}_\mathrm{s}>2$ [Figs.~\figref[f]{fig:pair_production}--\figref[j]{fig:pair_production}].
The transition therefore reflects a change in near-wake dynamics, not in nucleartion mechanisms.

As $\mathrm{Re}_\mathrm{s}$ increases, the pair-production rate and near-wake vortex density increase, strengthening vortex--vortex interactions and thus amplifying small asymmetries. 
This favors an alternating wake mode rather than symmetric dipolar emission, qualitatively analogous to the onset of the von K\'arm\'an instability in classical bluff-body flow, where a symmetric wake gives way to an alternating vortex street above a critical Reynolds number \cite{batchelor_2000}.
Consistently, symmetry breaking is already visible during pair production in Fig.~\figref[j]{fig:pair_production}, where the positive vortex separates more clearly than the negative one.
These observations suggest that the transition near $\mathrm{Re}_\mathrm{s}\approx2$ is the onset of a wake instability that drives the downstream reorganization of emitted vortex dipoles into same-sign clusters.

By contrast, for an impenetrable obstacle, vortices of opposite signs are nucleated on opposite sides of the obstacle, so the wake naturally develops an alternating, or V-shaped, arrangement of vortex dipoles even at low velocities \cite{sasaki_Phys.Rev.Lett._2010, li_NewJ.Phys._2019}. 
With the symmetry broken from the outset, clustering manifests as a smoother crossover in the force response~\cite{Reeves_PRA_2012}.


\section*{Appendix B: Dependence of $D_{\rm eff}$ on $v_0$}
\begin{figure}
    \centering
    \includegraphics[width=\linewidth]{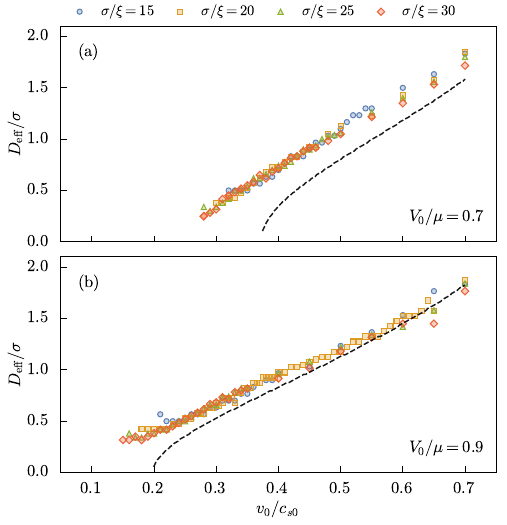}
    \caption{Flow-speed dependence of $D_\mathrm{eff}$. $D_{\mathrm{eff}}/\sigma$ as a function of $v_0/c_{\rm s0}$ for different Gaussian widths $\sigma$ at $V_0/\mu=0.7$ (a) and 0.9 (b). 
    Error bars are smaller than the symbols.
    Dashed lines show theoretical estimates from a perturbative steady-flow model using Eq.~\eqref{eq:potential_flow_result}.
    }
    \label{fig:D_eff_simul_theoretical}
\end{figure}

We consider steady flow past a Gaussian obstacle within a first-order perturbative potential-flow framework. In the stationary regime, and neglecting the quantum pressure term, the GPE takes a Bernoulli-like form,
\begin{subequations}\label{eq:Bernoulli}
\begin{align}
g\rho(\mathbf r) + V(\mathbf r) + \tfrac12 m v(\mathbf r)^2 &= g\rho_0 + \tfrac12 m v_0^2,\\[4pt]
\nabla\!\cdot[\rho(\mathbf r)v(\mathbf r)] &= 0. 
\end{align}
\end{subequations}
We examine small deviations around a uniform background flow,
\begin{subequations}\label{eq:perturbation}
\begin{align}
    \bm v(\mathbf r) &= -\,v_0\,\hat{\mathbf x} + \frac{\hbar}{m}\nabla\phi(\mathbf r),\\[4pt]
    \rho(\mathbf r)      &= \rho_0 + \delta\rho(\mathbf r),
\end{align}
\end{subequations}
with $|\frac{\hbar}{m}\nabla\phi|\ll v_0$ and $|\delta\rho|\ll\rho_0$. Substituting Eq.~\eqref{eq:perturbation} into Eq.~\eqref{eq:Bernoulli} and retaining only terms linear in $v_0/c_\mathrm{s0}$ yields the Poisson equation,
\begin{equation}
\partial_x^2 \phi(\mathbf r)
\;+\; \partial_y^2 \phi(\mathbf r)
\;=\; -\frac{v_0}{\hbar c_\mathrm{s0}^2}\,\partial_x V(\mathbf r).
\label{eq:linearized_Bernoulli}
\end{equation}
For a Gaussian obstacle
$V(\mathbf r)=V_0 \exp(-2r^2/\sigma^2)$, we use the polar coordinate $(r,\theta)$ defined as $x=r \cos{\theta}$ and $y=\sin{\theta}$. The solution for Eq.~\eqref{eq:linearized_Bernoulli} is given by 
\begin{equation}
\phi(\mathbf r)
= -V_0\,\frac{v_0}{4\hbar c_\mathrm{s0}^2}\,\sigma^2\,
\frac{\cos\theta}{r}\!
\left[1-\exp\!\Bigl(-\frac{2r^2}{\sigma^2}\Bigr)\right].
\label{eq:potential_flow_result}
\end{equation}

Combining Eq.~\eqref{eq:Bernoulli} and Eq.~\eqref{eq:perturbation} gives the local flow speed and density up to linear order in $v_0/c_\mathrm{s0}$,
\begin{equation}
\label{eq:linearized_Euler}
|\boldsymbol v|^2 = v_0^2 - \frac{2\hbar}{m}v_0\,\partial_x\phi,
\qquad
\rho = \rho_0 - \frac{V}{g} + \frac{\hbar}{g}v_0\,\partial_x\phi.
\end{equation}
Using Eq.~\eqref{eq:potential_flow_result}, we obtain
\begin{equation}
\frac{\hbar}{m}\partial_x\phi =
\frac{V_0}{4 m c_\mathrm{s0}^2}\,v_0\frac{\sigma^2}{r^2}\cos(2\theta).
\end{equation}
Thus, the flow speed is enhanced and the density is reduced on the lateral sides of the obstacle (\(\theta=\pm\pi/2\), \(\cos 2\theta=-1\)), while the opposite occurs along the flow axis (\(\theta=0 \,\text{or}\, \pi\), \(\cos 2\theta=1\)). Consequently, as \(v_0\) increases, the region satisfying the local supersonic condition, \(|\boldsymbol v(\boldsymbol r)|>c_\mathrm{s}(\boldsymbol{r})\), expands preferentially in the transverse direction. This leads to a lateral widening of the Mach-1 contour and thus an increase in the effective diameter \(D_{\mathrm{eff}}\).

In Fig.~\figref{fig:D_eff_simul_theoretical}, we compare the effective diameter obtained from this perturbative potential-flow solution with that extracted from full GPE simulations. 
The perturbative calculation captures, at a qualitative level, the increase of \(D_{\mathrm{eff}}\) with \(v_0\). 
It also predicts a minimum velocity above which the supersonic region acquires a finite extent; this threshold exceeds the actual critical velocity \(v_{\mathrm c}\). 
The discrepancy is attributed to the neglect of higher-order corrections and of the quantum-pressure term, which plays a crucial role in triggering vortex shedding.

\section*{Appendix C: Dipole emission frequency}

In the vortex dipole regime for $\mathrm{Re}_\mathrm{s} < 2$, the emitting frequency $f_\mathrm{D}$ shows a linear relationship with the flow speed $v_0$. 
This linearity was observed in previous experiments~\cite{lim_NewJ.Phys._2022, kwon_Phys.Rev.A_2015}, where a model of $f_\mathrm{D}=a (v_0-v_c)$ was suggested and employed to extract the critical velocity $v_c$. Furthermore, in Ref.~\cite{lim_NewJ.Phys._2022}, it was reported that the linear-response slope $a$ increased with an increase in the strength $V_0$ of the penetrable obstacle.

In Fig.~\figref[a]{fig:potential_vs_a}, we show our numerical results of $f_\mathrm{D}$ 
for different obstacle strengths with $\sigma/\xi = 20$.
The drag frequency is well described by the linear dependence on $v_0$, but the critical velocity is slightly higher than that determined by the linear extrapolation of the emitting frequency.
Noticeably, the value $a$ decreases with an increase of $V_0$, which is in stark contrast to the experimental observation.
We attribute this discrepancy to the finite-length driving protocol used in the experiments: the limited obstacle motion introduces acceleration phases and interactions between emitted vortices, which can affect the measured shedding frequency. 
Additional numerical simulations support this interpretation.

The relation $f_\mathrm{D}=a(v_0-v_c)$ suggests that the Strouhal number at the threshold velocity, $v_0=v_\mathrm{th}$, is given by 
\begin{equation}
    \mathrm{St}_{\mathrm{th}}
    = \frac{(f_{\mathrm{D}}/2) D_{\mathrm{eff}}}{ v_{\mathrm{th}}}
    = \frac{a(v_{\mathrm{th}} - v_c)D_{\mathrm{eff}}}{2v_{\mathrm{th}}}
    \approx \frac{\hbar}{m} \frac{a}{v_{\mathrm{th}}},
\end{equation}
where the last equality follows from $\mathrm{Re}_\mathrm{s}^\mathrm{th}\approx 2$.
The observation that both $v_{\mathrm{th}}$ and $\mathrm{St}_\mathrm{th}\sim\mathrm{St}_{0}$ decrease with an increase in $V_0$ is consistent with the dependence of $a$ on $V_0$.

\begin{figure}
    \centering
    \includegraphics[width=1\linewidth]{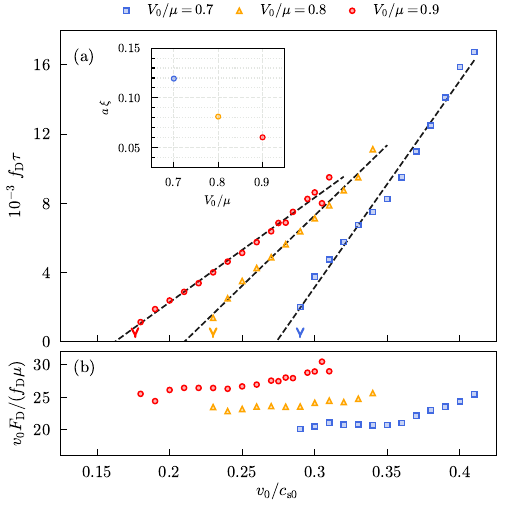}
    \caption{Periodic vortex-dipole emission. (a) Drag frequency $f_{\mathrm{D}}$ as a function of $v_0/c_{s0}$ for different values of $V_0/\mu$ with $\sigma/\xi = 20$. Dashed lines show the linear fits of $f_{\mathrm{D}} = a(v_0 - v'_c)$ to the data sets with $a$ and $v'_c$ being the fitting parameters.  
The critical velocities $v_c$ are indicated by color-matched wedge markers and are higher than $v'_c$.  
The inset shows the fitted values of $a$, normalized by $1/\xi$.
(b) Estimated vortex-dipole energy $E_\mathrm{v}=v_0 F_{\mathrm{D}}/f_{\mathrm{D}}$ [Eq.~(19)]. The energy is normalized by $\mu$.
}
    \label{fig:potential_vs_a}
\end{figure}

In addition, we consider the energetics of the obstacle's drag effect in the dipole emission regime. The work done by the drag force on the fluid should be equal to the dissipated energy through vortex dipole nucleation, which is expressed as
\begin{equation}
    F_{\mathrm{D}} v_0 = E_{\mathrm{v}} f_{\mathrm{D}},
\end{equation}
where $E_{\mathrm{v}}$ is the energy cost of generating a vortex dipole. 
In Fig.~\figref[b]{fig:potential_vs_a}, we show the variations of $E_\mathrm{v}=v_0 F_{\mathrm{D}}/f_{\mathrm{D}}$ for different obstacle conditions. $E_\mathrm{v}$ increases steadily with the background flow speed $v_0$.  
Since the energy of a pair of vortices of opposite-signs separated
by $D_{\mathrm{pair}}$ is given by $E_{\mathrm{v}} \propto \ln\!\left(D_{\mathrm{pair}}/\xi\right)$,
the increase in $E_v$ implies that the separation $D_{\mathrm{pair}}$ increases with the velocity.  
This behavior is fully consistent with the $v_0$-dependence of the effective obstacle diameter $D_{\mathrm{eff}}$ (Appendix B).

Together, the behaviors of $a(V_0)$, $v_0 F_{\mathrm{D}}/f_{\mathrm{D}}$, and $D_{\mathrm{eff}}(v_0)$ provide a coherent
description of the periodic vortex-dipole emission process.

\section*{Appendix D: Extension to different obstacles}
\begin{figure}[t]
    \centering
    \includegraphics[width=\linewidth]{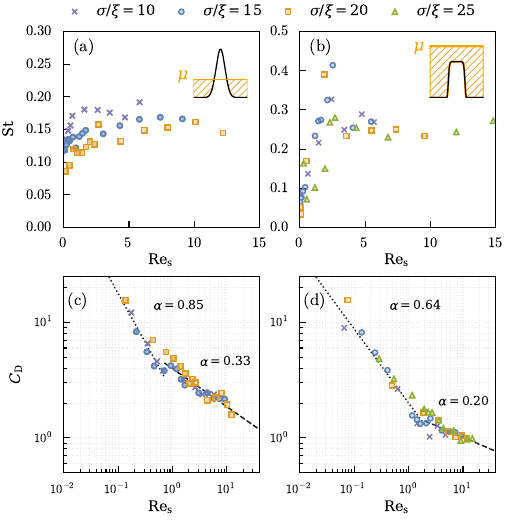}
    \caption{$\mathrm{St}$ and $C_\mathrm{D}$ versus $\mathrm{Re_s}$ for impenetrable Gaussian obstacles (a,c) and penetrable super-Gaussian obstacles (b,d). Insets in (a) and (b) show the corresponding obstacle profiles relative to the chemical potential $\mu$. 
    Dashed (dotted) lines in (c) show fits to 
$C_{\mathrm{D}} = b\,\mathrm{Re_s}^{-\alpha}$ for 
$\mathrm{Re_s} \ge 0.7$ ($\mathrm{Re_s} \le 0.7$). 
Similarly, the lines in (d) correspond to fits over 
$\mathrm{Re_s} \ge 2$ ($\mathrm{Re_s} \le 2$).
    }
    \label{fig:various_potential}
\end{figure}

To test whether the effective diameter $D_{\rm eff}$ provides a unified description beyond penetrable Gaussian obstacles, we repeated the $\mathrm{St}$ and $C_{\rm D}$ analysis using $D_{\rm eff}$ for two additional cases: an impenetrable Gaussian obstacle with $V_0/\mu=e$, and a penetrable super-Gaussian obstacle $V(\mathbf{r})=V_0 \exp{ [-2\left(r/\sigma\right)^6 ] }$ at $V_0/\mu=0.7$ (Fig.~\figref{fig:various_potential}). In both cases, $\textrm{St}$ and $C_{\rm D}$ collapse versus $\mathrm{Re_s}$ constructed with $D_{\rm  eff}$.

For the impenetrable Gaussian obstacle (Figs.~\figref[a]{fig:various_potential} and \figref[c]{fig:various_potential}), the shedding behavior and wake transition reproduce Ref.~\cite{reeves_Phys.Rev.Lett._2015}: a von Kármán street of same-sign vortex pairs (K2 regime) for $\mathrm{Re}_{\rm s}<0.7$, followed by irregular shedding beyond $\mathrm{Re}_{\rm s}\simeq 0.7$.
For a consistent comparison with Ref.~\cite{christenhusz_Phys.Rev.Lett._2025}, we analyzed the drag coefficient by subtracting the large-$\mathrm{Re_s}$ asymptote, $C_\mathrm{D0}$, rather than fitting $C_\text{D}$ directly.
This yields $C_\mathrm{D0}=2.06$, with power-law exponents $\alpha_1=1.37$ for $\mathrm{Re_s}<0.7$ and $\alpha_2=1.55$ for $\mathrm{Re_s}>0.7$. 
Compared with the results of $C_\mathrm{D0}=2.37$, $\alpha_1=1.06$ and $\alpha_2=0.72$ in Ref.~\cite{christenhusz_Phys.Rev.Lett._2025},
our exponents are larger, likely because we use the flow-defined $D_\mathrm{eff}$, which increases with an increase of $v_0$, rather than a fixed Thomas-Fermi zero-density size as the characteristic length.

For the penetrable super-Gaussian obstacle (Figs.~\figref[b]{fig:various_potential} and \figref[d]{fig:various_potential}), whose top is flatter than a standard Gaussian, the shedding dynamics closely mirror those of the penetrable Gaussian: 
periodic dipole emission at lower $\mathrm{Re_s}$ values, evolving into irregular shedding with same-sign vortex clusters at higher $\mathrm{Re_s}$ values. 
These results support $D_{\rm eff}$ as an obstacle-independent scale that organizes superfluid wake dynamics.


\begin{thebibliography}{99}
\bibitem{footnote_streetwidth}
Far downstream of the obstacle, the vorticity width was found to scale as $\sqrt{s}$, where $s$ is the distance from the obstacle, consistent with the 2D classical wake-width scaling~\cite{landau_1987}.
\end{thebibliography}
\end{document}